\documentstyle[epsfig]{article}        

\newcommand{\be}{\begin{equation}}
\newcommand{\ee}{\end{equation}}
\def\bea{\begin{eqnarray}}
\def\eea{\end{eqnarray}}

\def\qqg{$q {\bar q}g$}
\def\qq{$q {\bar q}$}

\input psfig

\newcommand{\gnufig}[3]{
\begin{figure}[t]
\begin{center}
{#3}
\end{center}
\caption{#2}
\label{fig:#1}
\end{figure}}

\newcommand{\Pomm}{{\tiny \mbox{I$\!$P}}}      

\begin{document}
\title{Soft Colour Interactions \\
and Diffractive DIS}
\author{W. Buchm\"uller\\
{\small\it DESY, Notkestrasse~85,~D-22603~Hamburg,~Germany}}
\date{}
\maketitle
\begin{abstract}
The basic ideas and some results of the semiclassical approach to diffractive 
DIS are briefly described. In the production of high-$p_{\perp}$ jets 
boson-gluon fusion is predicted to be the dominant partonic process. The 
$p_{\perp}$-spectrum and the two-jet invariant mass distribution provide a 
clear test of the underlying `hard' partonic process and the `soft' mechanism 
of colour neutralization.
\end{abstract}

The events with a large gap in rapidity in small-$x$ DIS \cite{wb:data} 
represent a puzzling phenomenon. The separation of a colour neutral cluster 
of `wee' partons from the proton, which then fragments independently of the 
proton remnant, is a non-perturbative, `soft' process. On the other hand, 
in rapidity gap events with high transverse momentum jets also a `hard' 
scattering process must take place. To disentangle the `soft' and
the `hard' aspects of `hard diffraction' is the main theoretical problem
of diffractive DIS \cite{wb:el}. 

Since diffractive processes are non-perturbative, a purely perturbative
approach, similar to ordinary parton model 
calculations, appears doomed to failure. In the following we shall describe
another attempt \cite{wb:bhm}, which is based on a high-energy expansion 
in the proton rest frame. At small $x$, the proton is treated as a classical
colour field localized whithin a sphere of radius $1/\Lambda$. Partonic
fluctuations of the virtual photon, \qq\, \qqg\ etc., are scattered by this 
colour field at high energies. Crucial ingredients of this semiclassical 
approach are light-cone techniques \cite{wb:bks} and the description of 
high-energy scattering processes in terms of Wilson lines \cite{wb:na}. 
A final state colour singlet partonic configuration is assumed to lead to
a diffractive event, since in this case the partonic cluster can fragment 
independently of the proton remnant. Correspondingly, a colour non-singlet 
partonic configuration yields an ordinary non-diffractive event.

Without any further ad-hoc assumptions, this simple physical picture leads 
to a number of predictions which are independent of the details of the
proton colour field, as long as it is soft with respect to the energies
of the incident partons. The following discussion is closely related to
Ref.~\cite{wb:bhm2}.\\

\noindent
{\large\bf Diffractive structure function}\\

\begin{figure}[ht]
\begin{center}
\vspace*{-.5cm}
\parbox[b]{11cm}{\psfig{width=10cm,file=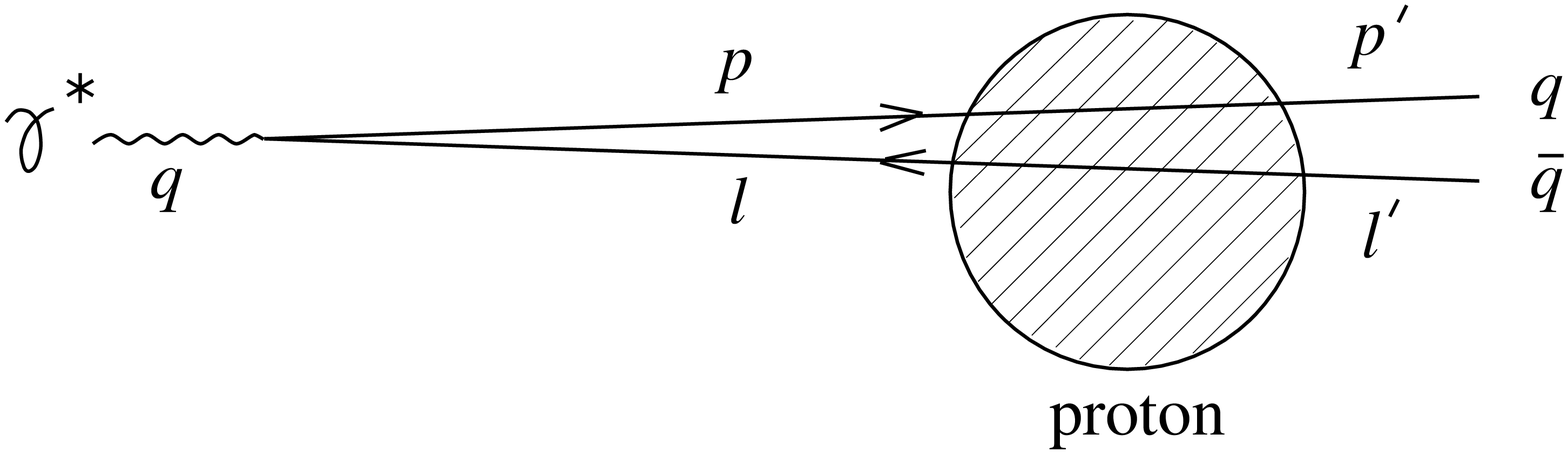}}\\
\end{center}
\refstepcounter{figure}
\label{f:wb:qq}
Figure \ref{f:wb:qq}: Exclusive two-jet production in the semiclassical 
approach.
\end{figure}

In the semiclassical approach inclusive and diffractive cross sections can 
be expressed in terms of a single non-perturbative quantity, 
$\mbox{tr}\,W^{\cal F}_{x_{\perp}}(y_{\perp})$, where 
\be\label{eq:wa}
W^{\cal F}_{x_\perp}(y_\perp)=U^\dagger(x_\perp+y_\perp)U(x_\perp)-1\,
\ee
is built from the non-Abelian eikonal factors $U$ and $U^\dagger$ of quark 
and antiquark whose light-like paths penetrate the colour field of the 
proton at transverse positions $x_\perp$ and $x_\perp+y_\perp$, respectively 
(cf.~Fig.~\ref{f:wb:qq}). The superscript ${\cal F}$  is used since quarks 
are colour triplets. As the colour field outside the proton vanishes 
$W^{\cal F}_{x_{\perp}}(y_{\perp})$ is essentially a closed Wilson loop 
through a section of the proton which measures an average of the proton 
colour field.

In an expansion in the transverse distance between quark and antiquark
one has
\be\label{eq:wsh}
\int_{x_{\perp}}\mbox{tr}W^{\cal F}_{x_{\perp}}(y_{\perp})
= - {1\over 4} y_{\perp}^2 C_1 + {\cal O}(y_{\perp}^4)\, .\label{intw}
\ee
The constant $C_1$ determines the variation of the inclusive structure 
function $F_2(x,Q^2)$ with $Q^2$ \cite{wb:bhm}. A comparison with  
boson-gluon fusion in the parton model shows the connection with the
gluon density,
\be\label{glue}
C_1 = 2 \pi^2 \alpha_s x G(x)\, .
\ee
Since $C_1$ is constant, $G(x) \sim 1/x$, 
which corresponds to a classical bremsstrahl spectrum of gluons.

Consider now the production of  \qq\ final states. Transverse momenta in the 
final state vary between $\Lambda$ and $Q$. Integration over this range
yields the dominant contribution to $F_2$ which is proportional to 
$\ln{Q/\Lambda}$. The inclusive structure function $F_2$ is linear in 
$\mbox{tr}W^{\cal F}_{x_{\perp}}(y_{\perp})$. In contrast, for diffractive
final states the structure function $F_2^D$ is quadratic in
$\mbox{tr}W^{\cal F}_{x_{\perp}}(y_{\perp})$ due to the projection on 
colour singlet \qq\ final states. Because of Eq.~(\ref{eq:wsh}), this
implies a suppression of large transverse momenta by one power of 
$l^2_{\perp}$. For kinematical reasons, the longitudinal momenta then have 
to be asymmetric, as in the aligned jet model \cite{wb:ajm}. With 
$\alpha= l_+/q_+ < 1/2$, one has
\be
l_{\perp} \sim \Lambda\, ,\qquad \alpha \sim {\Lambda^2\over Q^2}\ .
\ee 
For the diffractive structure function one obtains the result \cite{wb:bhm}
\be\label{pom}
F_2^D(x,Q^2,\xi) = {\beta\over \xi} \bar{F}(\beta)\, ,
\ee
where $\xi=x/\beta$ and $\beta=Q^2/(Q^2+M^2)$. $\bar{F}(\beta)$ can be 
expressed
as an integral over $\mbox{tr}W^{\cal F}_{x_{\perp}}(y_{\perp})$ and it is
therefore not calculable perturbatively. Eq.~(\ref{pom}) corresponds to
a pomeron structure function with $\alpha_{\Pomm}(0)=1$. \\

\noindent
{\large\bf Jets with large transverse momentum}\\ 

One can easily calculate the $p_{\perp}$-spectrum for  \qq\ final
states (cf.~Fig.~\ref{f:wb:qq}). The result reads \cite{wb:bhm2} 
\be\label{eq:ptqq}
\left.\frac{d\sigma_T}{d\alpha dp'^{2}_\perp dt}\right|_{t=0} \propto
C_1^2 \frac{(\alpha^2+(1-\alpha)^2)p'^{2}_\perp a^4}{(a^2+p'^{2}_\perp)^6}\, ,
\ee
where $t=(q-p'-l')^2$ is the momentum transfer to the proton and $a^2=\alpha
(1-\alpha)Q^2$. Since $C_1 \propto xG(x)$, the cross section is proportional
to the square of the gluon density. It is in fact identical to the result
obtained for two-gluon exchange in leading order \cite{wb:nn,wb:jb,wb:mw}.
The cross section integrated down to the transverse momentum $p'^2_{\perp}$
yields a contribution to the diffractive structure function $F_2^D$ which
is suppressed by $\Lambda^2/p'^2_{\perp,\mbox{\footnotesize cut}}$.

\begin{figure}[hb]
\begin{center}
\vspace*{-.5cm}
\parbox[b]{11cm}{\psfig{width=10cm,file=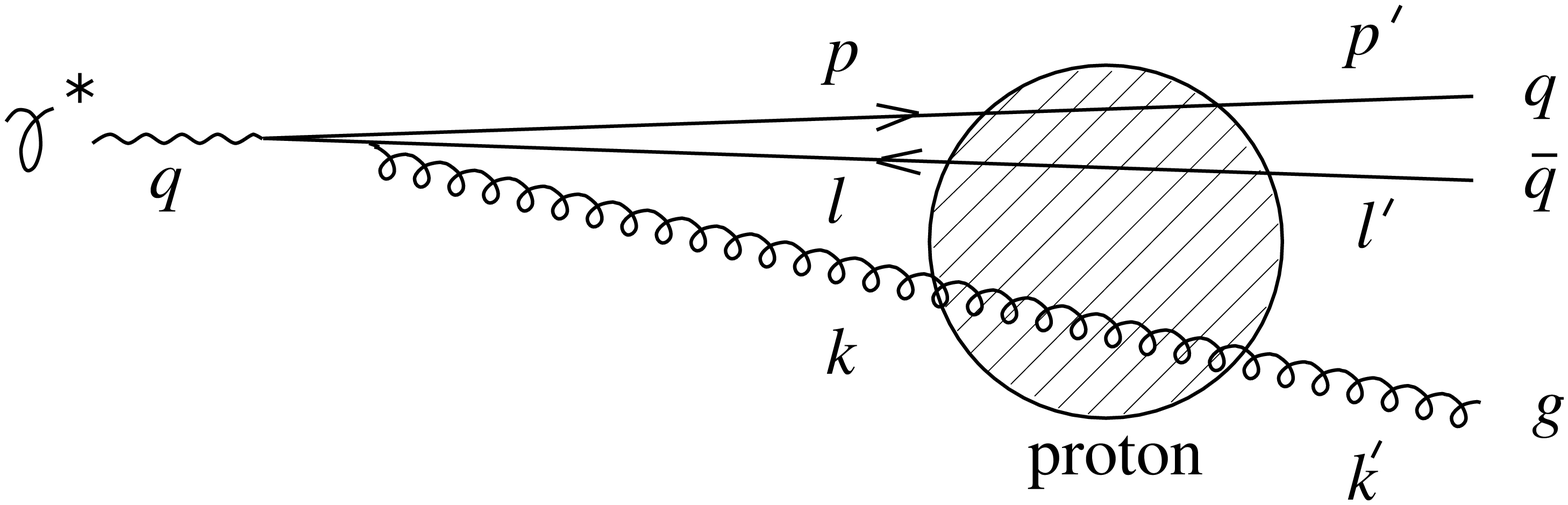}}\\
\end{center}
\refstepcounter{figure}
\label{f:wb:qqg}
Figure \ref{f:wb:qqg}: Two-jet production with an additional low transverse
momentum gluon.
\end{figure}

As shown in \cite{wb:bhm}, a `leading twist' contribution with jets of 
$p_{\perp}\sim Q$ requires at least three partons in the final state, one of 
which has low transverse momentum. It turns out that the dominant process has 
a low transverse momentum gluon (cf.~Fig.~\ref{f:wb:qqg}),
\be
k_{\perp} \sim \Lambda \, ,\qquad \alpha' \sim {\Lambda^2\over Q^2}\, 
\ee
and, correspondingly,
\be
k_+ = \alpha' q_+ \sim {\Lambda\over x}\, ,\qquad 
-k_-=-q_- +p_- +l_- \sim \Lambda x\, ,\qquad k^2 = - \Lambda^2\, .
\ee

\begin{figure}[hb]
\begin{center}
\vspace*{-.5cm}
\parbox[b]{10.5cm}{\psfig{width=9.5cm,file=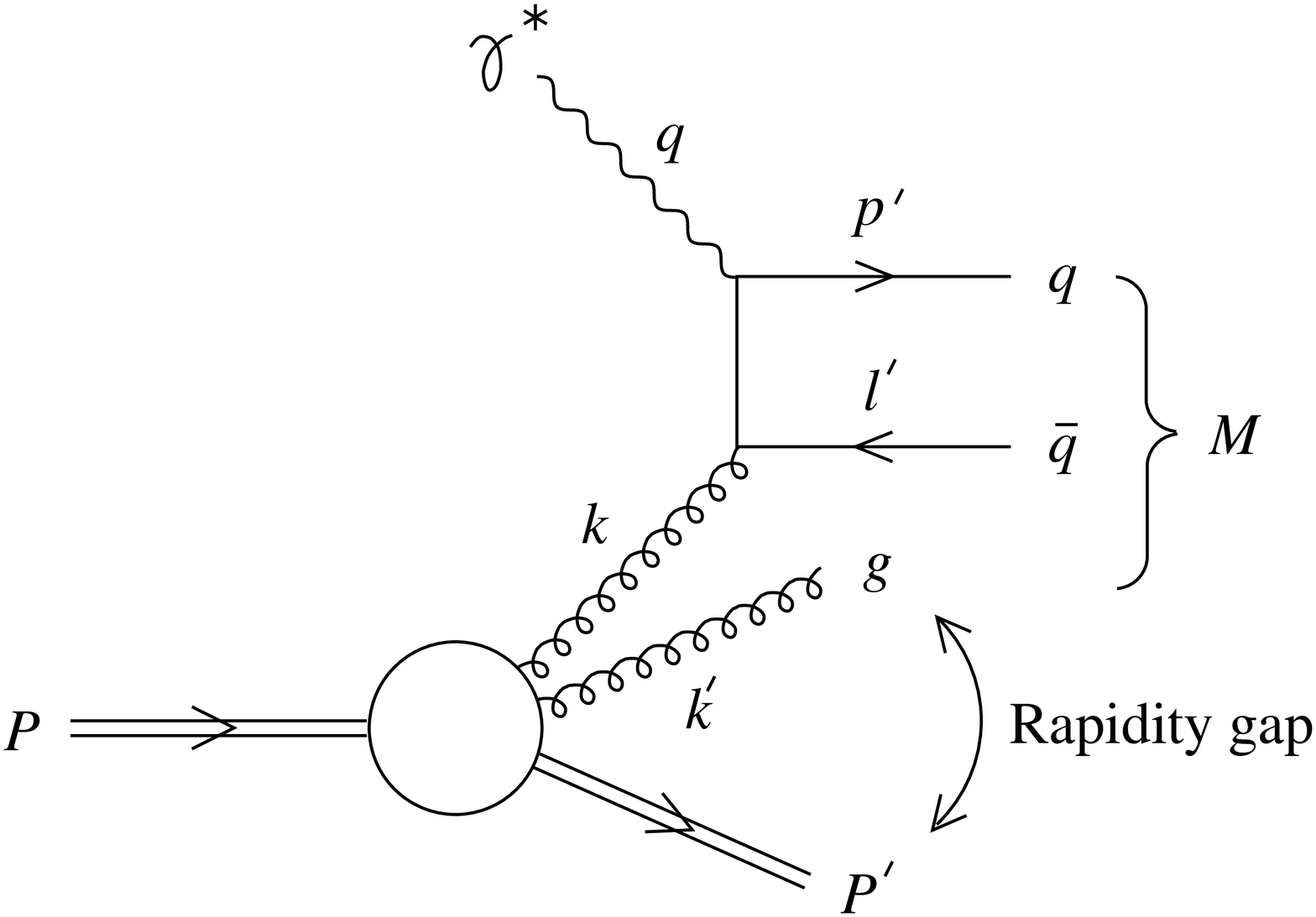}}\\
\end{center}
\refstepcounter{figure}
\label{f:wb:bgf}
Figure \ref{f:wb:bgf}: Interpretation of the process of Fig.~\ref{f:wb:qqg}
in terms of boson-gluon fusion in a frame where the proton is fast, e.g.,
the Breit frame.
\end{figure}

One may also view the various diffractive processes in a frame where the
proton is fast, e.g., the Breit frame. Note, that in the proton rest frame
$k_+\sim {\Lambda/x} \gg -k_-\sim \Lambda x$, whereas in the Breit frame
$-k_-\sim Q \gg k_+\sim {\Lambda^2/Q}$. The different cross sections can 
be written as
convolution of ordinary partonic cross sections with diffractive parton
densities \cite{wb:sop,wb:be,wb:ds,wb:ah}. The cross section for the process 
shown in Fig.~\ref{f:wb:qqg} then corresponds to boson-gluon fusion
(cf.~Fig.~\ref{f:wb:bgf}), 
\be\label{eq:bgfu}
{d\sigma_T\over d\xi dp'^2_{\perp}} = \int_x^\xi dy 
{d\hat{\sigma}_T^{\gamma^* g\rightarrow q\bar{q}}(y,p_{\perp}') 
 \over dp'^2_{\perp}} {dg(y,\xi)\over d\xi}\, .
\ee
The diffractive gluon density describes the probabilty to extract from the
proton a colour neutral pair of gluons: a virtual
gluon with momentum fraction $y$ which participates in boson-gluon fusion
and a real gluon with momentum fraction $\xi-y$ which contributes to the
diffractive final state. The diffractive gluon density is determined by
the proton colour field \cite{wb:ah},
\be\label{eq:dpd}
{dg(y,\xi)\over d\xi} = {1\over 8 y (\xi-y)}\int {d^2 
k'_\perp (k'_\perp)^2 \over (2\pi)^4}\int_{x_\perp}\left|\int {d^2 
k_\perp\over (2\pi)^2} {\mbox{tr}[\tilde{W}^{\cal A}_{x_\perp}(k'_\perp-
k_\perp)]t^{ij}\over k_\perp'^2 +k_\perp^2 u}\right|^2\, ,
\ee
where $t^{ij}$ is a tensor involving the transverse momenta $k_{\perp}$ and 
$k'_{\perp}$, and $u=(\xi-y)/y$. The function $\tilde{W}^{\cal A}_{x_\perp}$ 
is the Fourier transform of $W^{\cal A}_{x_\perp}$, which is defined as in 
Eq.~(\ref{eq:wa}), but with the $U$-matrices in the adjoint representation.
\gnufig{f:wb:psq}{The fraction of diffractive events with $p'^2_{\perp}$ above 
$p'^2_{\perp,\mbox{\footnotesize cut}}$ for $Q^2$ of 10 GeV$^2$ and 100
GeV$^2$ (lower and upper curve in each pair).}{
\setlength{\unitlength}{0.09bp}
\begin{picture}(4320,2592)(0,-100)
\includegraphics{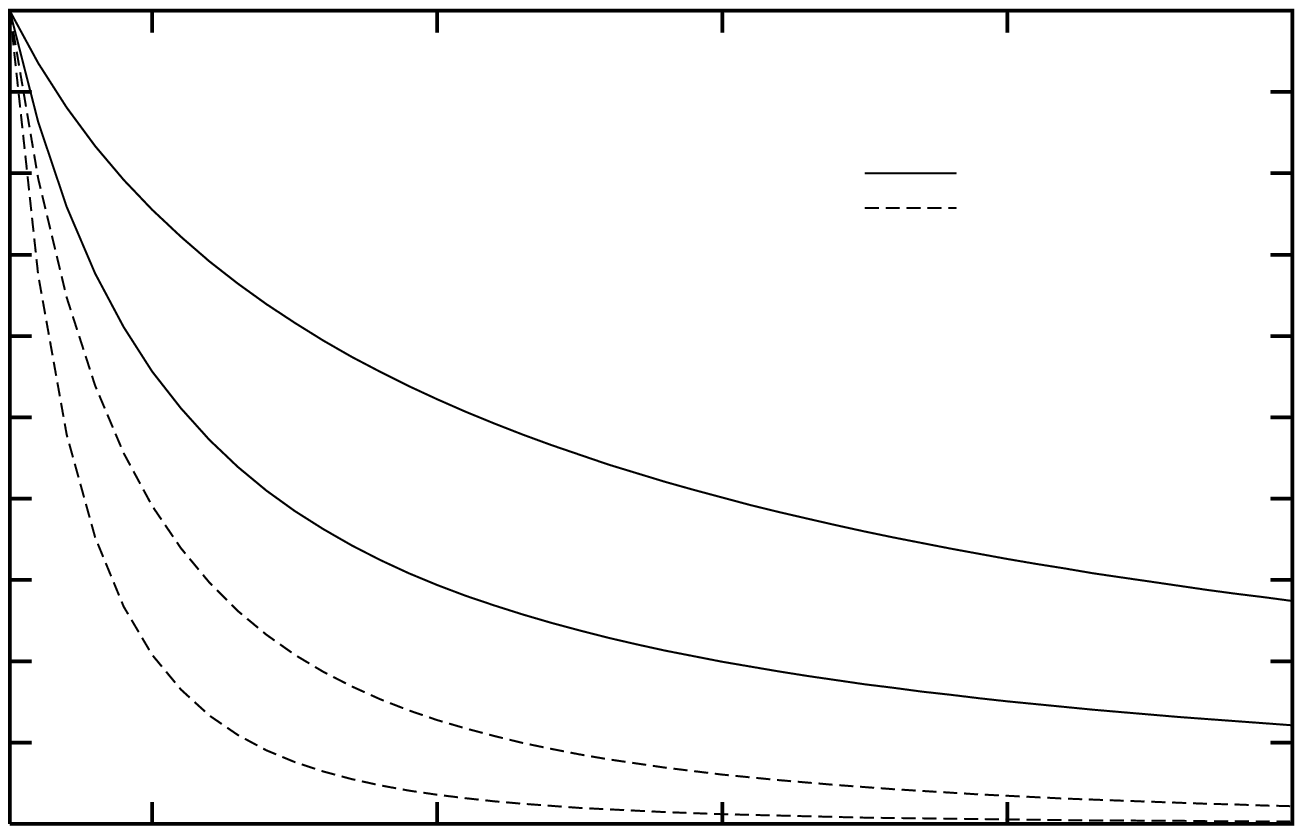}
\put(2876,1924){\makebox(0,0)[r]{ }}
\put(2876,2024){\makebox(0,0)[r]{ }}
\put(2679,1907){\makebox(0,0)[l]{$q \bar{q} ~~ $}}
\put(2679,2024){\makebox(0,0)[l]{$q \bar{q} g $}}
\put(2023,-118){\makebox(0,0)[l]{$p'^2_{\perp,\mbox{cut}} ~(\mbox{GeV}^2)$}}
\put(100,1321){%
\makebox(0,0)[b]{\shortstack{Fraction  of Events}}%
}
\put(4157,50){\makebox(0,0){50}}
\put(3336,50){\makebox(0,0){40}}
\put(2515,50){\makebox(0,0){30}}
\put(1694,50){\makebox(0,0){20}}
\put(873,50){\makebox(0,0){10}}
\put(413,2492){\makebox(0,0)[r]{1}}
\put(413,2258){\makebox(0,0)[r]{0.9}}
\put(413,2024){\makebox(0,0)[r]{0.8}}
\put(413,1789){\makebox(0,0)[r]{0.7}}
\put(413,1555){\makebox(0,0)[r]{0.6}}
\put(413,1321){\makebox(0,0)[r]{0.5}}
\put(413,1087){\makebox(0,0)[r]{0.4}}
\put(413,853){\makebox(0,0)[r]{0.3}}
\put(413,618){\makebox(0,0)[r]{0.2}}
\put(413,384){\makebox(0,0)[r]{0.1}}
\put(413,150){\makebox(0,0)[r]{0}}
\end{picture}
}

From Eqs.~(\ref{eq:bgfu}) and (\ref{eq:dpd}) one obtains for the differential 
cross section in the leading-$\ln(1/x)$ approximation,
\be\label{eq:ptqqg}
\frac{d\sigma_T}{d\alpha dp_\perp'^2} \propto
\alpha_s\frac{(\alpha^2+(1\!-\!\alpha)^2)\,(p_\perp'^4+a^4)}
{(a^2+p_\perp'^2)^4}\,\ln(1/x) \, .
\ee
A comparison of Eqs.~(\ref{eq:ptqq}) and (\ref{eq:ptqqg}) shows that the
$p_{\perp}$-spectrum for the \qqg\ configuration is much harder than that
for the \qq\ configuration. This is expected since in boson-gluon fusion 
$p_\perp$ is distributed  logarithmically between $\Lambda$ and $Q$, thus 
resulting in a significant high-$p_\perp$ tail. The quantitative differences 
are particularly  pronounced in the integrated cross section with a lower cut 
on transverse  momentum $p'^2_{\perp,\mbox{\footnotesize cut}}$. The shape of
the momentum distribution is shown in Fig.~4. Each curve is 
normalized to its value at 
$p'^2_{\perp,\mbox{\footnotesize cut}} = 5 $~GeV$^2$. 

The \qq\ and \qqg\ final states also differ with respect to the invariant
mass distribution of the two jets. The additional wee gluon contributes
significantly to the diffractive mass of the final state \cite{wb:bhm2}.
Rather similar to the diffractive production of high $p_{\perp}$-jets are
the qualitative features of diffractive open charm production \cite{wb:mfd}.\\

\noindent
{\large\bf Comparison with other approaches}\\

It is instructive to compare the described results with those of other
approaches to diffractive DIS. The phenomenology of the semiclassical
approach is qualitatively very similar to {\bf `soft' pomeron models}
\cite{wb:data,wb:ds}, if the pomeron is `gluonic'. In leading order it 
corresponds to $\alpha_{\Pomm}(0)=1$. The projection on colour singlet final 
states for diffractive DIS yields asymmetric parton configuations in the proton
rest frame, which is the basis of the {\bf aligned jet model} \cite{wb:ajm}. 

A qualitative difference with respect to {\bf `hard' pomeron models}, so far 
mostly two-gluon exchange \cite{wb:nn,wb:jb,wb:mw,wb:ds}, is the 
$p_{\perp}$-spectrum. It will be interesting to see how important \qqg\ final 
states are in the two-gluon exchange model. In the semiclassical approach 
the dominant partonic 
process is boson-gluon fusion, as in the {\bf boson-gluon fusion model} for 
diffraction \cite{wb:bh}. However, in contrast to this model, the 
semiclassical approach predicts an additional low transverse momentum gluon
in the final state. Similar in spirit is the {\bf soft colour interaction
model} \cite{wb:gi}, where some soft non-perturbative gluon exchange is
incorporated in a Monte Carlo event generator.
 
Finally, important questions which remain to be studied in the semiclassical
approach are: the evolution in $Q^2$, the effect of higher order corrections
on the $\xi$-dependence, the treatment of further low transverse momentum
gluons in the diffractive final state and, on the theoretical side, the
question of the validity of the semiclassical approximation.


\end{document}